# Dynamic formation of spherical voids crossing linear defects

Y. A. Bioud[1,2], M. Rondeau[1], A. Boucherif[1], G. Patriarche[3], D. Drouin[1] and R. Arès[1]

[1] *Laboratoire Nanotechnologies Nanosystèmes (LN2) – CNRS UMI-3463, Institut Interdisciplinaire d'Innovation Technologique (3IT), Université de Sherbrooke, 3000 Boulevard Université, Sherbrooke, J1K OA5 Québec, Canada*
[2] *Institut quantique, Département de physique, Université de Sherbrooke, Sherbrooke, J1K 2R1, Québec, Canada*
[3] *Centre de Nanosciences et de Nanotechnologies Université Paris-Saclay C2N – Site de Marcoussis Route de Nozay, F-91460 Marcoussis, France*

Corresponding authors: y.bioud@usherbrooke.ca, richard.ares@usherbrooke.ca.

A predictive model for the evolution of porous Ge layer upon thermal treatment is reported. We represent an idealized etched dislocation core as an axially symmetric elongated hole and computed its dynamics during annealing. Numerical simulations of the shape change of a completely spherical void via surface diffusion have been performed. Simulations and experiments show individual large spherical voids, aligned along the dislocation core. The creation of voids could facilitate interactions between dislocations, enabling the dislocation network to change its connectivity in a way that facilitates the subsequent annihilation of dislocation segments. This confirms that thermally activated processes such as state diffusion of porous materials provide mechanisms whereby the defects are removed or arranged in configurations of lower energy. This model is intended to be indicative, and more detailed experimental characterization of process parameters such as annealing temperature and time, and could estimate the annealing time for given temperatures, or vice versa, with the right parameters.

Recently, the microstructure transformation innovative concept derived from the modification of nano/microcavities in crystalline germanium during high temperature annealing led to the development of novel technologies such as germanium on nothing membranes [1]–[3] and voided Ge/Si virtual substrates [4]–[12]. Such processes rely on the idea of removing bulk material, and creating patterned or randomly distributed, free space within the semiconductor crystal. Resulting voided material augments the number of surface atoms. Such atoms have less tying bonds and therefore can become mobile when the crystal temperature rises in an annealing process. At high temperature, it was shown that moving surface atoms create diffusing currents that reshape the overall morphology of the introduced voids [13]. Neglecting evaporation, matter conservation implies that void volume remains constant during its evolution to an equilibrium state. Nichols et al. have proposed a dynamic model describing the morphological changes of a surface of revolution due to capillarity-induced surface diffusion assuming the isotropy of surface tension and surface self-diffusion coefficients [14]. Ghannam et al. gave a physical interpretation for the transformation of cylindrical macropores embedded in crystalline silicon during high temperature treatments [13]. Starting from these frameworks, we represent an idealized etched dislocation core as an axially symmetric elongated hole and computed its dynamics during annealing.

### Chemical potential and curvature

The difference in chemical potential Δμ between a flat (no curvature) and a curved surface on a solid is given by,

$$\Delta\mu = \gamma C_m V_{at} \quad (1)$$

where $V_{at}$ is the volume of one atom, γ is the surface tension of the solid and $C_M$ is the mean curvature of the surface defined in [14] as,

$$C_M = C_T + C_\theta \quad (2)$$

where $C_T$ and $C_\theta$ are two perpendicular curvatures on a surface parametrized in 3D space. Equation 1 can be derived using the Ostwald-Freundlich equation relating vapor pressure and curvature and the Gibbs-Helmholtz equation for change in chemical potential with pressure [15].

If the curvature varies along an arc *s* of the surface then atoms drift towards lower chemical potentials with a drift velocity υ along this arc obtained using the Nernst-Einstein relation,

$$v = -\frac{D_s}{k_B T}\frac{\delta \mu}{\delta s} \quad (3)$$

where $D_s$ is the surface diffusion constant of the atoms, T the temperature, $k_B$ the Boltzmann constant and $-\frac{\delta \mu}{\delta s}$ is the driving force for a single parameter parameterized surface. If the area density of particles is given by σ, using equations 1 and 3 the surface current is given by equation 4,

$$J = -\frac{\gamma V_{at} \sigma D_s}{k_B T}\frac{\delta C_m}{\delta_s} \quad (4)$$

## Morphology dynamics

Fig. 1a and 1b illustrate Mullins [14] rationale to derive the equation for the dynamics of the void morphology by using matter conservation.

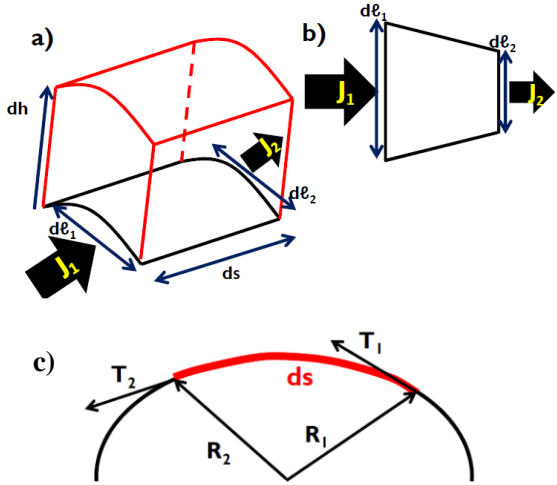

Figure 1. (a) Growth of δh small surface patch δlδs (b) Top view of (a). (c) Visualization of the unit tangent vector change $T_1$, $T_2$ on the surface when traveling a distance d*s*.

The volume variation at one point along an arc *s* of the surface is necessarily due to the increase or decrease of atoms taken or left by a surface current *J*. The number of atoms carried by *J* is given by *Jδlδs*, where *δl* is a profile length element and *δt* a time increment. If a current $J_1$ enters a surface patch $\delta s(\delta l_1 + \delta l_2)/2$ and a current $J_2$ comes out, then the difference in number of atoms gives $(J_1\delta l_1 - J_2\delta l_2)\delta t$, each with a small atomic volume $V_{at}$. This volume change is translated to a change in height h that enters in a small prism volume change $\delta h \delta s (\delta l_1 + \delta l_2)/2$. This conservation law gives the change in height,

$$(J_1 \delta l_1 - J_2 \delta l_2)\delta t V_{at} \sim \delta s \delta h \frac{(\delta l_1 + \delta l_2)}{2} \quad (5)$$

When $J_2 \rightarrow J_1$, $\delta l_1 \rightarrow \delta l_2$, then the left-hand side of equation 5 reads -δ(Jδl) and the average length is $(\delta l_1 + \delta l_2)/2 = \delta l$,

$$-\delta(J\delta l)\delta t V_{at} \sim \delta s \delta h \delta l \quad (6)$$

The change in height δh depends on time and the position δs the void profile. Rearranging equation 6 we find what looks like a differential equation in time and space.

$$\frac{\delta h}{\delta t} = -\frac{V_{at}}{\delta l}\frac{\delta(J\delta l)}{\delta s} \quad (7)$$

When substituting equation 4 in 7 it yields the same result than Mullins [14].

$$\frac{\delta h}{\delta t} = \frac{\gamma V_{at}^2 \sigma D_s}{K_B T \delta l}\frac{\delta}{\delta s}\left(\delta l \frac{\delta C_M}{\delta s}\right) \quad (8)$$

Equation 8 relates local mean curvature $C_M$ to morphology change.
Curvature is defined as the variation of a unit tangent vector T about an arc s on the surface, see equation 9. Referencing to figure 1c, if $T_1$ varies to appreciably different unit vector $T_2$ on an incremental distance d*s*, then it is said that the curvature is important.

$$C = \left\|\frac{dT}{ds}\right\| \quad (9)$$

This definition is reasonable since it is expected that the tangent vector to a surface changes "rapidly" its orientation when the surface is highly curved. The curvature may be signed, if the tangent vector turns counter clockwise with the normal sticking out the surface then it is positive C > 0, otherwise it is negative C < 0.

## Development for axially symmetric shapes

When the void space is axially symmetric as illustrated in figure 2, then the radius ρ in the xy plane can be simply parameterized with z, ρ(z) and changes with time t, ρ (z,t). In the left hand side of figure 2 the change in ρ (z,t) is illustrated with circles of various radius.

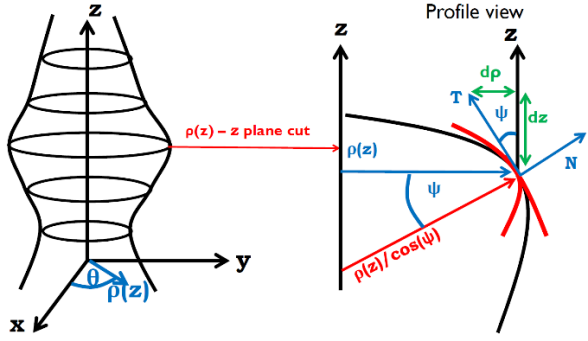

Figure 2. Axially symmetric rippled hole about the z axis. The circle are cuts in the x-y plane and the curve in the right-hand side is a cut in the ρ-z plane. The vector T is tangent to the profile surface towards positive z and θ is the angle ρ makes with the x axis. The vector $\hat{\theta}$ is also tangent to the surface but perpendicular to T. Vectorial product between T and $\hat{\theta}$ yields a vector normal to the surface N. The angle that T makes with the z axis is noted ψ. The red curves under the profile view are the local perpendicular curvatures.

For this symmetry δl becomes δl = ρ(z,t)dθ where θ is defined in figure 2. Rewriting, equation 8,

$$\frac{\delta h}{\delta t} = B \frac{1}{\rho(z,t)} \frac{\delta}{\delta s}\left(\rho(z,t)\frac{\delta C_M}{\delta s}\right) \quad (10)$$

where $B = \gamma V_{at}^2 \delta D_s / k_B T$. The expression for the mean curvature $C_M$ can be derived for axial symmetry.

### Calculation of mean curvature for axial symmetry

The idea now is to compute the local mean curvature, equation 8, using equation 9 considering an axial symmetry. For this, cylindrical coordinates are used and the local unit vectors are noted $\hat{\rho}$, $\hat{\theta}$, and $\hat{z}$. Noting the Cartesian basis unit vector $\hat{x}$, $\hat{y}$ and $\hat{z}$, the cylindrical basis can be written,

$$\hat{\rho} = cos(\theta)\hat{x} + cos(\theta)\hat{y},$$
$$\hat{\theta} = -sin(\theta)\hat{x} + cos(\theta)\hat{y} \quad (11)$$

The position vector r, illustrated by the green arrow in the left-hand side of figure 2 can therefore be written,

$$r = \rho(z,t)\hat{\rho} + z\hat{z} \quad (12)$$

Knowing the definitions 11 the first tangent vector T used to compute the mean curvature 2 is expressed using the surface tangent angle with z, noted ψ, which is limited to ψ ∈ [0; π] and written as,

$$T = sin(\psi)\hat{\rho} + cos(\psi)\hat{z} \quad (13)$$

An arc length ds following the tangent T is given by,

$$ds_T = \sqrt{dz^2 + d\rho^2} \quad (14)$$

From these definitions and equation 9, the first term of the mean curvature is obtained

$$C_T = \left|\frac{dT}{ds_T}\right| = \frac{d\psi}{ds_T} \quad (15)$$

The second tangent vector to the surface and normal to T is $-\hat{\theta}$. The elemental arc length about that direction is computed using the length of the red arrow in the right-hand side of figure 2 multiplied by dθ,

$$ds_\theta = \frac{\rho(z,t)}{cos(\psi)}d\theta \quad (16)$$

Still computed with equation 9 the curvature about $-\hat{\theta}$ is,

$$C_\theta = -\left|\frac{d\hat{\theta}}{ds_\theta}\right| = -\frac{cos(\psi)}{\rho(z,t)} \quad (17)$$

The minus sign for rotation orientation of $-\hat{\theta}$. Total mean curvature is the sum of equations 15 and 17,

$$C_M = \frac{d\psi}{ds_T} - \frac{cos(\psi)}{\rho(z,t)}$$

With the expression of mean curvature in hand, it is now possible to express equation 10 in cylindrical coordinates with a single variable, z. From chain rule derivation and using equation 14, the derivative with respect to profile arc is given by equation 19,

$$\frac{\delta}{\delta s} \rightarrow \frac{1}{\sqrt{1+(\rho_z)^2}}\frac{\delta}{\delta z} \quad (19)$$

where $\rho_z = d\rho/dz$. Replacing equations 18 and 19 in equation 10,

$$\frac{\delta h}{\delta t} = B\frac{1}{\rho(z,t)}\frac{1}{\sqrt{1+(\rho_z)^2}}\frac{\delta}{\delta z}\left(\rho(z,t)\frac{1}{\sqrt{1+(\rho_z)^2}}\frac{\delta}{\delta z}\left(\frac{1}{\sqrt{1+(\rho_z)^2}}\frac{\delta \psi}{\delta z}\right.\right.$$
$$\left.\left. - \frac{cos(\psi)}{\rho(z,t)}\right)\right) \quad (20)$$

Knowing that ψ = tan⁻¹(ρ_z) the derivative with z is readily evaluated as $\frac{\delta \psi}{\delta z} = \frac{\rho_{zz}}{(1+\rho_z^2)}$ where ρ_zz is the second derivative of ρ(z,t) with respect to z. Furthermore, $\cos(\psi) = 1/\sqrt{1+\rho_z^2}$ from equation 14. Finally, δh/δt is the height variation perpendicular to the surface and oriented inwards since this is for a void, $\delta h/\delta t = -N\, \delta r/\delta t$. The product $-N\, \delta r/\delta t$ is given by $\delta h/\delta t = -1/\sqrt{1+\rho_z^2}\,\delta\rho(z,t)/\delta t$ equation 20 simplifies to,

$$\frac{\delta\rho(z,t)}{\delta t} = \frac{B}{\rho(z,t)}\frac{\delta}{\delta z}\left(\frac{\rho(z,t)}{\sqrt{1+(\rho_z)^2}}\frac{\delta}{\delta z}\left(\frac{1}{\rho(z,t)\sqrt{1+(\rho_z)^2}} - \frac{\rho_{zz}}{(1+\rho_z^2)^{\frac{3}{2}}}\right)\right) \quad (21)$$

Equation 21 is a fourth order differential expression, the same as the one expressed by Coleman et al. [16]. Solving this equation requires a numerical approach in order to obtain dynamic behavior of the void morphology.

## Numerical solutions

One way to solve equation 21 numerically is simply using finite differences to express the spatial derivatives and the fourth-order Runge-Kutta method to drive the time integration.

The following simulation was carried using centered finite difference scheme for spatial derivatives,

$$\frac{d\rho}{dz} \sim \frac{\rho(z_{i+1}) - \rho(z_{i-1})}{z_{i+1} - z_{i-1}} \quad (22)$$

$$\frac{d^2\rho}{dz^2} \sim \frac{\rho(z_{i+1}) - 2\rho(z_i) + \rho(z_{i-1})}{(z_{i+1} - z_i)^2} \quad (23)$$

$$\frac{dC_M}{d_z} \sim \frac{C_M(z_{i+1}) - C_M(z_{i-1})}{z_{i+1} - z_{i-1}} \quad (24)$$

$$\frac{d^2C_M}{dz^2} \sim \frac{C_M(z_{+1}) - 2C_M(z_i) + C_M(z_{i-1})}{(z_{i+1} - z_i)^2} \quad (25)$$

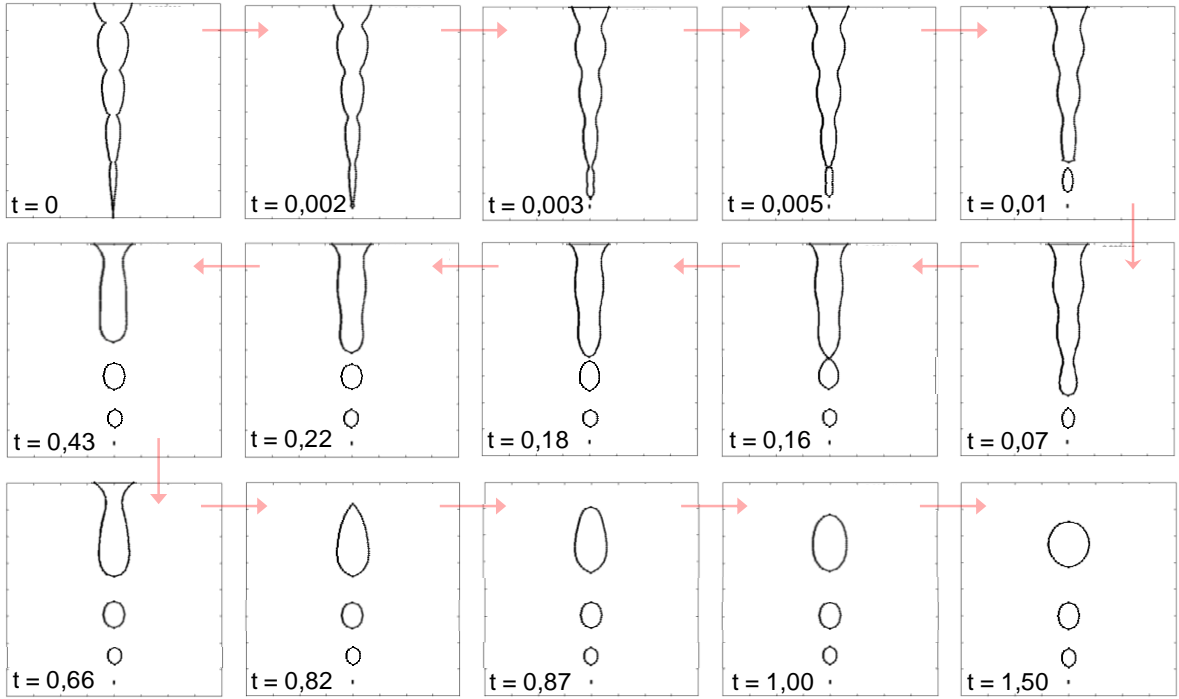

Figure 3. Shape evolution of an axially symmetric rippled elongated hole obtained by numerical simulation based on the model described above.

With,

$$C_M = \frac{d^2\rho}{dz^2} \frac{1}{\left(1 + \left(\frac{d\rho}{dz}\right)\right)^{3/2}} - \frac{1}{\rho(z,t)\sqrt{1 + \left(\frac{d\rho}{dz}\right)^2}} \quad (26)$$

Equation 20 is transformed into a unit-less expression by setting time and spatial units at B = 1. Figure 3 shows, frame by frame, the calculated change in the rippled elongated hole. The initial profile is constructed so to be similar to the etched pores in the bulk material, which are locally axially symmetric. The mean aspect ratio is height/radius = 16 and the initial ripples along the profile is the product between a cone of half-angle 2.5° with successive circles equations of increasing $z$ center points. The border conditions are set so that there is always matter coming from far away; therefore, the maximum $z$ point never changes radius $\rho$ ($\bar{z}_{Max}$) value. The profile is divided in 100 spatial elements with a height set to $\bar{z}_{Max}$= 8, $d\bar{z}$ = 8/100, where the top bar indicates that it is a unit-less quantity. The time integration is carried out using the GNU Scientific

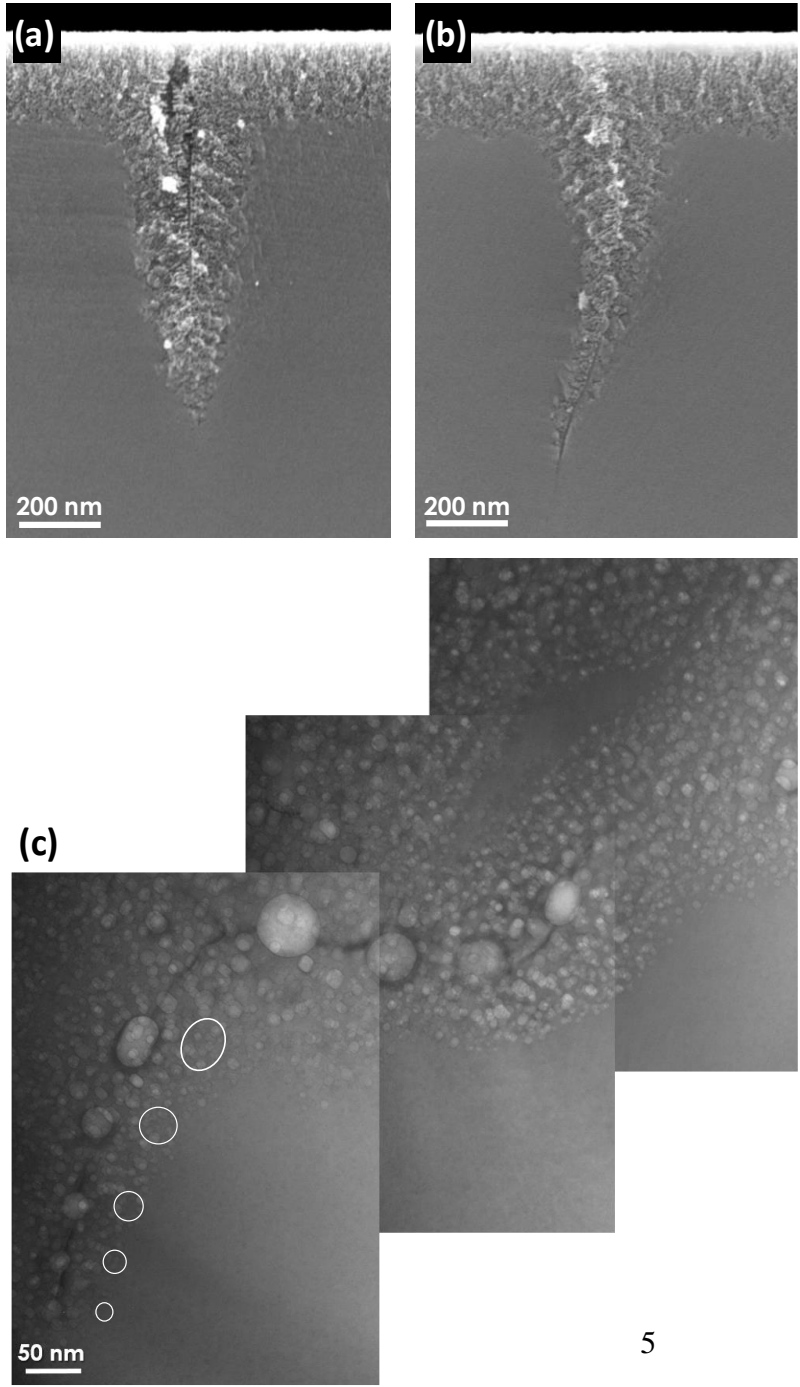

Figure 4: Different morphologies of "deep-etched dislocation shape" crossing threading dislocation cores obtained by electrochemical etching of Ge/Si layer at 1.5 mA/cm² (a,b). TEM image shows porous shape spheroidizes into aligned and separate large spherical voids with increasing size (c). The pinning of a dislocation segment in the array of nanovoids is observed.

Library's (GSL) fourth-order Runge-Kutta ODE solver with adaptive time-step.
First, the hole shrinks laterally near the opening via emergence of an overhang, whereas it expands near the bottom. The shrinking finally causes a topological change, forming a closed void. Next, relaxation of the void shape occurs and eventually the shape becomes spherical. The same operation happens simultaneously to form aligned voids. The spheroidization time increases with increasing size of voids having a larger diffusion surface. Neglecting evaporation, matter conservation implies that void volume remains constant during its evolution to an equilibrium state.

### Experimental observation

Experimentally, the Ge porous structure, obtained by bipolar electrochemical etching [17]–[20], is composed of highly interconnected lateral mesopores, separated by a primary vertical, nearly axially symmetric pore that follows the orientation of dislocations. The resulting morphology, so-called "deep-etched dislocation shape", is presented in the Figure 4(a) and 4(b). At an annealing temperature of 550°C, a significant morphological transformation occurred, in which the porous structure self-organized into a network of nanovoids following the dislocation lines, as shown in the TEM microphotograph of Figure 4(c). The porous shape evolves into aligned spheres and separate large spherical voids with increasing size, as predicted from the simulations. We notice also the presence of other smaller neighbouring voids, eventually coalesce by increasing annealing duration obeying the theory describing Oswald ripening phenomenon [13], [21].

### Conclusion

We studied the mechanism of void formation by observing the change in shape of high-aspect-ratio holes on Ge surfaces during annealing. We compared the morphological transformation of the "deep-etched dislocation shape" obtained experimentally with the surface-diffusion driven "rippled elongated" shape generated by numerical simulations. We found that the extremely strong morphological instability arises near hole openings, causing spontaneous closure of the hole and forming aligned and separate large spherical voids.